# Quantum Electrodynamics of a Nanocavity Coupled with Exciton Complexes in a Quantum Dot

Makoto Yamaguchi, Takashi Asano, Kazunobu Kojima and Susumu Noda

*Department of Electronic Science and Engineering, Kyoto University,*

*Katsura, Nishikyo-ku, Kyoto 615-8510, Japan*

## ABSTRACT

Here, a comprehensive theory of the couplings between a nanocavity and exciton complexes in a quantum dot is developed, which successfully predicts the spectral triplet in the strong coupling regime that has been observed in several experiments but is unexpected according to conventional cavity quantum electrodynamics. The quantum anti-Zeno effect is found to play an essential role in the appearance of the central peak in the triplet under a low-excitation regime. The effect of hyperfine interactions is also discussed, which results in the cavity-mediated mixing of bright and dark exciton states. These results provide significant insights into solid-state cavity quantum electrodynamics.

# I. INTRODUCTION

The coupled system of a nanocavity and a quantum dot (QD) has been extensively investigated in the past few years because it has promising properties for quantum information processing (QIP) [1], single photon sources [2], and ultimately low-threshold lasers [3]. In this system, the following two phenomena were expected to occur naturally, analogous to the atomic systems in cavity quantum electrodynamics (QED): first, light emission only at the QD transition energy under off-resonant conditions; and second, vacuum Rabi splitting (VRS) under on-resonant conditions. However, experimental reports have shown the following clear deviations from these features: first, light emission from the cavity occurs even when it is largely detuned from the QD [4–8]; and second, spectral triplets are formed by additional bare cavity lines between the VRS lines under on-resonant conditions [4,7,8]. These features are unique to semiconductor systems and are not predicted by conventional cavity QED in atomic systems. Recently, we reported that the quantum anti-Zeno effect (AZE), induced by pure-dephasing in the QD, plays a key role in the off-resonant cavity light emission [9,10]. In our previous study, the QD was modeled by a simple two-level system (TLS), and the effect of pure-dephasing (or non-referred measurements) was explained as follows: it broadens the transition energy of the TLS, which makes it possible for the tail of the transition energy to overlap with the cavity energy, and generates an additional way in which to interact with the cavity. Because of this interaction, a cavity photon is created with the transition of the excited TLS to the ground state, which is generally known as the AZE. This effect seems negligible in typical cases, because the tail of the TLS transition energy at the cavity energy is so weak that the interaction is not expected to be sufficiently large to overcome the direct spontaneous emission from the TLS to free space. However, the unique situation achieved in semiconductor nanocavity QED systems strongly enhances the AZE, resulting in the off-resonant cavity light emission, as follows: first, the coupling constant between the cavity and the TLS is huge due to the small cavity volume; second, the quality factor of the cavity is relatively large; and third, direct spontaneous emission from the TLS to free space is strongly suppressed due to the in-plane photonic band-gap effect in the case of two-dimensional photonic crystal (2DPC) nanocavities [11] shown in Fig. 1(a). These features are well described by a factor, $F$, introduced in Ref. [10], which expresses the ratio of the integrated intensity at the cavity energy to that at the TLS energy as $F$: $(1 - F)$. Here, $F$ is given by

$$F \approx \frac{\Gamma_{spon} + \gamma_{phase}}{\Gamma_{spon}(\delta\omega_{TLS,cav}/g_{TLS,cav})^2 + \Gamma_{cav} + (\Gamma_{spon} + \gamma_{phase})}, \quad (1)$$

where $2\Gamma_{spon}$ is the direct spontaneous emission rate of the TLS to free space, $2\Gamma_{cav}$ is the optical damping rate of the cavity, $2\gamma_{phase}$ is the pure-dephasing rate of the TLS, $\delta\omega_{TLS,cav}$ is the detuning of the TLS from the cavity, and $g_{TLS,cav}$ is the coupling constant between the TLS and the cavity.

Thus, the off-resonant cavity light emission is successfully explained by this nanocavity-enhanced AZE, and its essential formula is also derived. An equivalent analysis was also reported by another group independently [12]. However, the whole emission mechanism including the on-resonant spectral triplet remains a puzzling problem. Furthermore, recent experiments [7] revealed unique couplings of the nanocavity with individual charging states in the QD, and found that dark excitons (DX[0]) are optically activated by the cavity-mediated mixing with bright excitons (BX[0]). These features are beyond the TLS model of the QD, and are not explained by any previously reported theories. It is important to clarify these emission mechanisms, because they directly influence the design and performance of relevant applications.

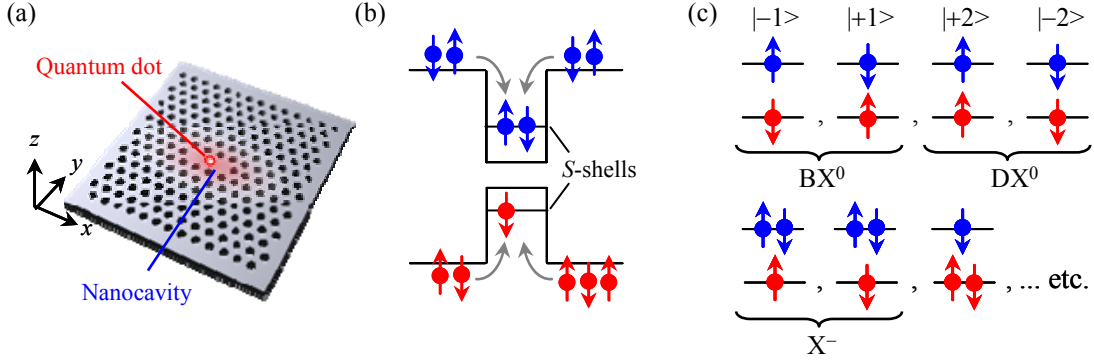

FIG. 1 (color). Schematic of the system. (a) QD embedded in a 2DPC nanocavity. Here, the $x$, $y$, and $z$ directions correspond to the [110], [1−10], and [001] directions, respectively. (b) First confined levels ($S$-shell) in the QD. (c) Examples of different charge configurations in the $S$-shells where blue and red arrows represent the electron spins and heavy-hole pseudo-spins, respectively. Individual states in $BX^0$ and $DX^0$ are labeled by the total angular momentum of the carriers.

In this paper, we develop a comprehensive theory of the couplings between a nanocavity and a QD based on the quantum master equation (QME), in which individual charge configurations are taken into account (Fig. 1(b) and (c)). The numerically calculated results not only reproduce the unique features of the QD exciton complexes but also predict the on-resonant spectral triplet successfully. In this system, the AZE is shown to play an essential role in the appearance of the central cavity peak of the spectral triplet in collaboration with the charging states under a low-excitation regime. Furthermore, the cavity-mediated mixing of bright and dark exciton states is also shown to occur as a result of hyperfine (HF) interactions enhanced by the nanocavity.

A theoretical model that describes the dynamics of a nanocavity and a QD in the framework of the QME is presented in Section II of the present paper. Numerically calculated results that include emission lines of individual charge configurations, and a discussion of the origin of the spectral triplet and the cavity-mediated mixing, are presented in Section III. Finally, our results and perspectives are summarized in Section IV. Further details are presented in the appendices at the end of the paper.

## II. THEORETICAL METHODS

In all previous studies treating the spectral features of a single QD coupled with a nanocavity, to the best of our knowledge, various types of exciton complex in the QD were neglected [9–12,13]. By contrast, here we formulate a theory of exciton complexes in the QD coupled with a nanocavity, by taking into account individual charge configurations and their Coulomb and exchange interactions.

In our treatment, a Stranski-Krastanov indium arsenide (InAs) QD grown on a gallium arsenide (GaAs) substrate is assumed, where the $x$ and $y$ direction are defined as [110] and [1–10], respectively. As shown in Fig. 1(b), electrons with $S_z = \pm 1/2$ spins and heavy holes with $J_z = \pm 3/2$ pseudo-spins are independently injected into the first confined levels ($S$-shells) in the QD, and various exciton states are formed, as shown in Fig. 1(c). The quantum dynamics in this system can be described by the QME in the interaction picture, that is,

$$\mathrm{d}\rho_S(t)/\mathrm{d}t = -i\hbar[H_S(t), \rho_S(t)] + L\rho_S(t), \tag{2}$$

which is focused on the subject system consisting of the cavity mode and the $S$-shells in the QD. Here, $\rho_S(t)$ is a reduced density operator, $H_S(t)$ denotes interactions inside the subject system (non-Markov processes), and the Liouvillian $L$ denotes irreversible processes (Markov processes). The basis of the states is then taken as $|i_{1/2}, j_{-1/2}\rangle \otimes |k_{3/2}, l_{-3/2}\rangle \otimes |p_{\mathrm{cav}}\rangle$ ( $i, j, k, l = 0$ or $1$ , $p = 0, 1, 2, ...$ ), where $|i_{1/2}, j_{-1/2}\rangle$ denotes $i$ ($j$) electrons with spin +1/2 (−1/2), $|k_{3/2}, l_{-3/2}\rangle$ denotes $k$ ($l$) heavy holes with pseudo-spin +3/2 (−3/2), and $|p_{cav}\rangle$ corresponds to $p$ photons in the cavity. As a result, a total of 16 different charging states are taken into account, including not only $BX^0$ and $DX^0$ but also bi-excitons ($XX^0$), positively ($X^+$) and negatively ($X^-$) charged excitons, and other optically inactive states. Hereafter, for descriptive convenience, each $BX^0$ state of $|0_{1/2}, 1_{-1/2}\rangle \otimes |1_{3/2}, 0_{-3/2}\rangle$ and $|1_{1/2}, 0_{-1/2}\rangle \otimes |0_{3/2}, 1_{-3/2}\rangle$ is labeled by $|+1\rangle$ and $|-1\rangle$, and each $DX^0$ state of $|1_{1/2}, 0_{-1/2}\rangle \otimes |1_{3/2}, 0_{-3/2}\rangle$ and $|0_{1/2}, 1_{-1/2}\rangle \otimes |0_{3/2}, 1_{-3/2}\rangle$ is denoted by $|+2\rangle$ and $|-2\rangle$, as shown in Fig. 1(c).

### A. Non-Markov processes

Initially, we explain non-Markov processes expressed by

$$H_S(t) = H_{CL} + H_{EX} + H_{LM}(t). \tag{3}$$

Here, $H_{CL}$ and $H_{EX}$ denote the Coulomb and exchange interactions of carriers in the $S$-shells, respectively, and $H_{LM}(t)$ denotes the light-matter couplings between the cavity and excitons. $H_{CL}$ in Eq. (3) is expressed by

$$H_{CL} = J^{e\text{-}e} n_\downarrow n_\uparrow + J^{h\text{-}h} m_\downarrow m_\uparrow - J^{e\text{-}h}(n_\downarrow + n_\uparrow)(m_\downarrow + m_\uparrow), \tag{4}$$

where $n_\sigma \equiv c_\sigma^\dagger c_\sigma$ and $m_\sigma \equiv d_\sigma^\dagger d_\sigma$ ($\sigma = \uparrow$ or $\downarrow$) are the number operators of the electrons and holes defined by the Fermi creation and annihilation operators of the electrons ($c_\sigma^\dagger$, $c_\sigma$) and holes ($d_\sigma^\dagger$, $d_\sigma$). The first term in Eq. (4) describes the Coulomb repulsion energy between the two electrons with different spins characterized by $J^{e\text{-}e}$. In a similar way, the second and third terms also represent the Coulomb charging energies between different carriers characterized by $J^{h\text{-}h}$ and $J^{e\text{-}h}$. By contrast, $H_{EX}$ in Eq. (3) is described by

$$H_{\text{EX}} = -2\delta_0 s_h^z s_e^z + \delta_1 \left(s_h^+ s_e^- + s_h^- s_e^+\right)/2 + \delta_2 \left(s_h^+ s_e^+ + s_h^- s_e^-\right)/2, \tag{5}$$

where $s_e^+ \equiv c_\uparrow^\dagger c_\downarrow$, $s_e^- \equiv c_\downarrow^\dagger c_\uparrow$, and $s_e^z \equiv (n_\uparrow - n_\downarrow)/2$ are the spin operators of electrons, and $s_h^+ \equiv d_\uparrow^\dagger d_\downarrow$, $s_h^- \equiv d_\downarrow^\dagger d_\uparrow$, and $s_h^z \equiv (m_\uparrow - m_\downarrow)/2$ are the effective spin operators [14] corresponding to the two heavy-hole states $J_z = \pm 3/2$. Here, $\delta_0$, $\delta_1$, and $\delta_2$ are the parameters determined by the anisotropy of the QD [14–16]. $\delta_0$ determines the energy difference between the $|\pm 1\rangle$ and $|\pm 2\rangle$ states. By contrast, $\delta_1$ determines the fine-structure splitting (FSS) of the $|\pm 1\rangle$ states (BX$^0$) and their eigen-states are mixed into $(|+1\rangle \pm |-1\rangle)/\sqrt{2}$. These modified bright states $(|+1\rangle - |-1\rangle)/\sqrt{2}$ and $(|+1\rangle + |-1\rangle)/\sqrt{2}$ are known to interact with the $x$ and $y$ polarized light fields, respectively. In the same way, $\delta_2$ determines the FSS of the $|\pm 2\rangle$ states (DX$^0$) and their eigen-states turn into $(|+2\rangle \pm |-2\rangle)/\sqrt{2}$. Finally, $H_{\text{LM}}(t)$ is formulated by $H_{\text{LM}}(t) = H_{\text{LM}}^{+1}(t) + H_{\text{LM}}^{-1}(t)$ with

$$H_{\text{LM}}^{+1}(t) = \hbar c_\downarrow^\dagger d_\uparrow^\dagger a_{\text{cav}} g_{+1} \exp(i\delta\omega t) + \text{H.c.}, \tag{6}$$

$$H_{\text{LM}}^{-1}(t) = \hbar c_\uparrow^\dagger d_\downarrow^\dagger a_{\text{cav}} g_{-1} \exp(i\delta\omega t) + \text{H.c.}, \tag{7}$$

where $a_{\text{cav}}$ is the Bose annihilation operator of the cavity mode, and $\delta\omega$ is the detuning of the QD transition frequency from the cavity. Note that the two coupling constants, $g_{\pm 1}$, both depend on the polarization of the cavity at the QD and are mutually related. Given the respective Bloch functions of the conduction and heavy-hole bands, $g_{\pm 1}$ can be written as

$$g_{\pm 1} = \mp g(\cos\theta_{\text{cav}} \mp i\sin\theta_{\text{cav}}), \tag{8}$$

where $\theta_{\text{cav}}$ is the angle between the cavity polarization and the $x$ axis. The magnitude of the interaction between the cavity and each optically active state of BX$^0$, X$^+$, X$^-$, and XX$^0$ is consistently determined by the relation. In order to understand this, it might be instructive to note that $H_{\text{LM}}(t)$ can be rewritten as

$H_{\text{LM}}(t) = H_{\text{LM}}^x(t) + H_{\text{LM}}^y(t)$ with

$$H_{\text{LM}}^x(t) = \hbar\sqrt{2}\cos\theta_{\text{cav}} g \hat{B}_x^\dagger \hat{a}_{\text{cav}} \exp(i\delta\omega t) + \text{H.c.}, \quad (9)$$

$$H_{\text{LM}}^y(t) = i\hbar\sqrt{2}\sin\theta_{\text{cav}} g \hat{B}_y^\dagger \hat{a}_{\text{cav}} \exp(i\delta\omega t) + \text{H.c.}, \quad (10)$$

where $B_x$ and $B_y$ are defined as $(d_\downarrow c_\uparrow - d_\uparrow c_\downarrow)/\sqrt{2}$ and $(d_\downarrow c_\uparrow + d_\uparrow c_\downarrow)/\sqrt{2}$, respectively. Considering an example of $B_x^\dagger|\text{GS}\rangle = -(|+1\rangle - |-1\rangle)/\sqrt{2}$ with $|\text{GS}\rangle \equiv |0_{1/2}, 0_{-1/2}\rangle \otimes |0_{3/2}, 0_{-3/2}\rangle$, it can be understood that $B_x$ ($B_x^\dagger$) represents the annihilation (creation) operator of the $x$-polarized BX$^0$ state. A similar principle applies to $B_y$ and $B_y^\dagger$. Thus, the magnitude of the interaction between the cavity and each optically active state is consistently determined by the relation of Eq. (8).

### B. Markov processes

Next, we discuss the Markov processes expressed by the Liouvillian, $L$, in Eq. (2), which can be divided into four terms:

$$L = L_{\text{cav}} + L_{\text{spon}} + L_{\text{phase}} + L_{\text{inj}}. \quad (11)$$

Here, the first term denotes the photon escape from the cavity to free space at a rate of $2\Gamma_{\text{cav}}$, the second term denotes the direct photon emission from the QD to free space at a rate of $2\Gamma_{\text{spon}}$, the third term denotes the pure-dephasing of the electrons at a rate of $2\gamma_{\text{phase}}^{\text{e}}$ and heavy holes at a rate of $2\gamma_{\text{phase}}^{\text{h}}$, and the fourth term indicates that the electrons and holes are independently injected at an equal rate of $2P$ with random spins (pseudo-spins). Here, the differences in each rate between the carriers and spins are neglected for simplicity. Each Liouvillian can be written in the Lindblad form as

$$L_{\text{cav}} X = -\Gamma_{\text{cav}}\left(a_{\text{cav}}^\dagger a_{\text{cav}} X + X a_{\text{cav}}^\dagger a_{\text{cav}} - 2 a_{\text{cav}} X a_{\text{cav}}^\dagger\right), \quad (12)$$

$$L_{\text{spon}} X = -\Gamma_{\text{spon}} \sum_\sigma \left(c_\sigma^\dagger d_{-\sigma}^\dagger d_{-\sigma} c_\sigma X + X c_\sigma^\dagger d_{-\sigma}^\dagger d_{-\sigma} c_\sigma - 2 d_{-\sigma} c_\sigma X c_\sigma^\dagger d_{-\sigma}^\dagger\right), \quad (13)$$

$$L_{\text{phase}} X = -\gamma^{\text{e}}_{\text{phase}} \sum_\sigma (n_\sigma X + X n_\sigma - 2 n_\sigma X n_\sigma) - \gamma^{\text{h}}_{\text{phase}} \sum_\sigma (m_\sigma X + X m_\sigma - 2 m_\sigma X m_\sigma), \quad (14)$$

$$L_{\text{inj}} X = -P \sum_\sigma (c_\sigma c_\sigma^\dagger X + X c_\sigma c_\sigma^\dagger - 2 c_\sigma^\dagger X c_\sigma) - P \sum_\sigma (d_\sigma d_\sigma^\dagger X + X d_\sigma d_\sigma^\dagger - 2 d_\sigma^\dagger X d_\sigma), \quad (15)$$

where $-\sigma$ denotes the inversion of $\sigma$, and the summation of $\sigma$ is taken for $\uparrow$ and $\downarrow$. $L_{\text{cav}}$ is the same as the commonly used Liouvillian [9,10,13]. By contrast, $L_{\text{spon}}$ and $L_{\text{phase}}$ are an extension of our previous work [10]. It should be noted that these Liouvillians of $L_{\text{spon}}$ and $L_{\text{phase}}$ can consistently express each Markov process of various charging states at the same time. Here, the original form of $L_{\text{phase}}$ was initially derived as a result of Coulomb interactions with surrounding carriers [10], but phonon-mediated pure-dephasing also takes the same form under the Markov approximation, because the most important point for the derivation of the $L_{\text{phase}}$ is that the effective interaction Hamiltonian between the carriers of the QD and the environmental degrees of freedom commutes with the number operators of the carriers in the QD. By contrast, the pumping model proposed by Laussy *et al.* [13] is expanded for $L_{\text{inj}}$ in order to describe the independent injection of electrons and holes.

### C. Emission spectra

Finally, we show the expression of the emitted photon spectrum $S(\omega)$ in the steady state, which can be divided into two terms:

$$S(\omega) = S_{\text{cav}}(\omega) + S_{\text{spon}}(\omega). \quad (16)$$

Here, $S_{\text{cav}}(\omega)$ and $S_{\text{spon}}(\omega)$ denote the spectral components following the spatial radiation pattern of the cavity mode and the QD, respectively. Each spectrum is characterized by the operator that produces the corresponding Lindblad form in the Liouvillians $L_{\text{cav}}$ and $L_{\text{spon}}$. Therefore, $S_{\text{cav}}(\omega)$ is described as

$$S_{\text{cav}}(\omega) = \frac{2\Gamma_{\text{cav}}}{\pi} \text{Re}\left[\int_0^\infty d\tau \langle a_{\text{cav}}^\dagger(t) a_{\text{cav}}(t+\tau)\rangle \exp(i\omega\tau - \gamma_{\text{reso}}\tau)\right], \tag{17}$$

where the spectral resolution is defined by $2\gamma_{\text{reso}}$. By contrast, the Liouvillian $L_{\text{spon}}$ in Eq. (13) can be rewritten by $B_x$ and $B_y$ as

$$L_{\text{spon}} X = -\Gamma_{\text{spon}}\left(B_x^\dagger B_x X + X B_x^\dagger B_x - 2 B_x X B_x^\dagger\right) - \Gamma_{\text{spon}}\left(B_y^\dagger B_y X + X B_y^\dagger B_y - 2 B_y X B_y^\dagger\right), \tag{18}$$

where two Lindblad forms are included. Therefore, $S_{\text{spon}}(\omega)$ is expressed by a sum of two linearly-polarized components:

$$S_{\text{spon}}(\omega) = S_x(\omega) + S_y(\omega), \tag{19}$$

with

$$S_x(\omega) \equiv \frac{2\Gamma_{\text{spon}}}{\pi} \text{Re}\left[\int_0^\infty d\tau \langle B_x^\dagger(t) B_x(t+\tau)\rangle \exp(i\omega\tau - \gamma_{\text{reso}}\tau)\right], \tag{20}$$

$$S_y(\omega) \equiv \frac{2\Gamma_{\text{spon}}}{\pi} \text{Re}\left[\int_0^\infty d\tau \langle B_y^\dagger(t) B_y(t+\tau)\rangle \exp(i\omega\tau - \gamma_{\text{reso}}\tau)\right]. \tag{21}$$

Here, $S_x(\omega)$ and $S_y(\omega)$ denote the x-polarized and y-polarized components in $S_{\text{spon}}(\omega)$. By contrast, $S_{\text{spon}}(\omega)$ can also be divided into two circularly-polarized components as expected from the original form of $L_{\text{spon}}$ in Eq. (13). These expressions are described in Appendix A.

## III. RESULTS AND DISCUSSION

We have calculated the steady-state emitted photon spectra of $S(\omega)$ by transforming the QME into differential equations of matrix elements for $\rho_S(t)$ with the quantum regression theorem [17,18]. Unless otherwise stated, the parameters listed in Table I are used here. $J^{e-e}$, $J^{h-h}$, and $J^{e-h}$ are estimated from the calculated values reported in Ref. [19], and the other parameters are consistent with the experimental results [4,7,8] where the cavity mode is $y$ polarized at the QD position ($\theta_{cav} = \pi/2$). Here, the $2\gamma_{phase}$ is defined as $2\gamma_{phase}^{e} + 2\gamma_{phase}^{h}$, and it is assumed that $2\gamma_{phase}^{e} = 2\gamma_{phase}^{h}$. Notably, a sufficiently low carrier-injection rate is used in the calculation in order not to saturate the states in the QD.

| Quantity | Value | Unit | Quantity | Value | Unit |
|---|---|---|---|---|---|
| $J^{e-e}$ | $2.6 \times 10^1$ | meV | $2\hbar\Gamma_{cav}$ | $6.9 \times 10^1$ | μeV |
| $J^{h-h}$ | $3.0 \times 10^1$ | meV | $2\hbar\Gamma_{spon}$ | $4.4 \times 10^1$ | neV |
| $J^{e-h}$ | $2.9 \times 10^1$ | meV | $2\hbar\gamma_{phase}$ | $3.0 \times 10^1$ | μeV |
| $\delta_0$ | $2.5 \times 10^2$ | μeV | $2\hbar P$ | $3.3 \times 10^1$ | neV |
| $\delta_1$ | $-3.0 \times 10^1$ | μeV | $2\hbar g$ | $2.1 \times 10^2$ | μeV |
| $\delta_2$ | $-1.0 \times 10^1$ | μeV | $\theta_{cav}$ | $\pi/2$ | Rad |

TABLE I. Standard parameters used in this work.

## A. Spectral triplets

Fig. 2(a) and (b) show the calculated $S(\omega)$ values with changes of the detuning $\delta\omega$ in two different cases where the pure-dephasing is either ignored ($2\gamma_{phase} = 0$ μeV) or taken in consideration ($2\gamma_{phase} = 30$ μeV), respectively. As shown in Fig. 2(a), there are basically four emission lines of $X^+$, $X^-$, $BX^0$, and $XX^0$. Here, the $X^+$ and $X^-$ lines are both composed of two oppositely circularly-polarized lines, whereas the $BX^0$ and $XX^0$ lines include two non-degenerated linearly-polarized lines in the $x$ and $y$ directions. When the cavity is tuned to $X^+$, it splits into two lines with a separation of $2\hbar g = 2.1 \times 10^2$ μeV, corresponding to VRS. As the tuning proceeds, VRS can again be seen in resonance with $BX^0$ (the splitting is of $2\sqrt{2}\hbar g = 3.0 \times 10^2$ μeV). In comparison with $X^+$, there is a line in $BX^0$ that is unaffected by the cavity. This corresponds to the $x$-polarized state in $BX^0$, which cannot interact with the $y$-polarized cavity field. Furthermore, when the cavity is tuned to $BX^0$, a simultaneous splitting of $XX^0$ occurs. This

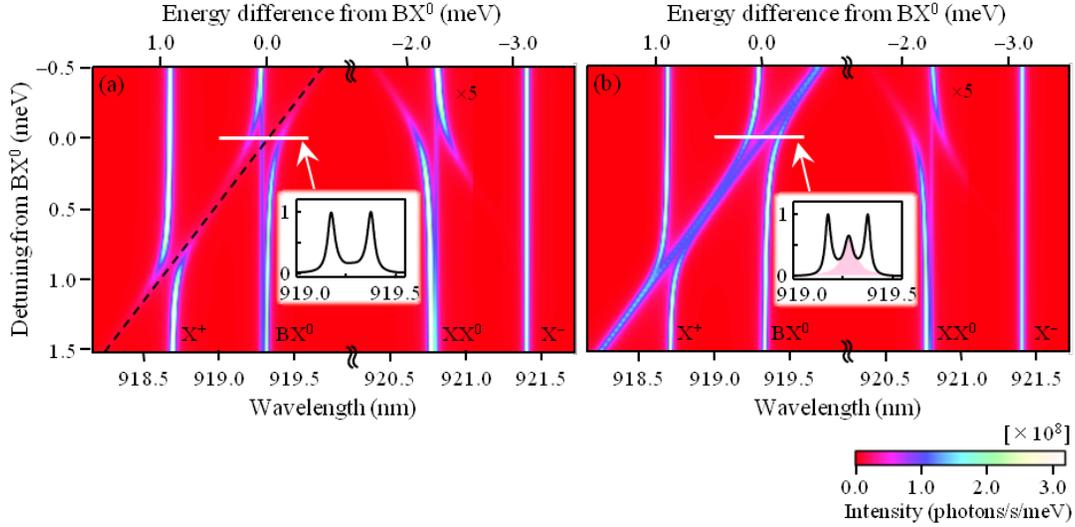

FIG. 2 (color). Calculated spectra (a) with and (b) without pure-dephasing. The dashed line represents the resonant wavelength of the cavity. The spectral resolution is set as $2\gamma_{reso} = 30$ μeV in the former and $2\gamma_{reso} = 0$ μeV in the latter to compensate for the difference of line broadening. The color scale around the $XX^0$ lines (920.5–920.0 nm) has been offset by a factor of five. The insets show the normalized spectra of VRS without the *x*-polarized components when the detuning from $BX^0$ is zero.

is because the final states of the $XX^0$ transitions are determined by $BX^0$. Again, the unaffected line in $XX^0$ is the *x*-polarized component. These features are consistent with the experimental results [7], indicating that our treatment of the exciton complexes works well; however, the result shown in Fig. 2(a) does not reproduce the off-resonant cavity light emission and the on-resonant triplet (the *x*-polarized line in $BX^0$ is ruled out as an origin of the triplet because the central peak in the triplet preserves the exact wavelength, line width, and polarization of the cavity mode in the experiments). By contrast, the emission spectra are drastically changed, as shown in Fig. 2(b), when the pure-dephasing is taken into account. Even if the cavity is detuned from all emission lines, the cavity light emission appears to be in agreement with previous reports [9–12]. Moreover, beyond our expectations, bare cavity lines constantly appear between the VRS peaks when the cavity is tuned to $X^+$ and $BX^0$ (see the insets to Fig. 2). These features have also been observed experimentally [4,7,8]. Our results indicate that both the AZE and the practical model of a QD are necessary to reproduce the spectral triplet.

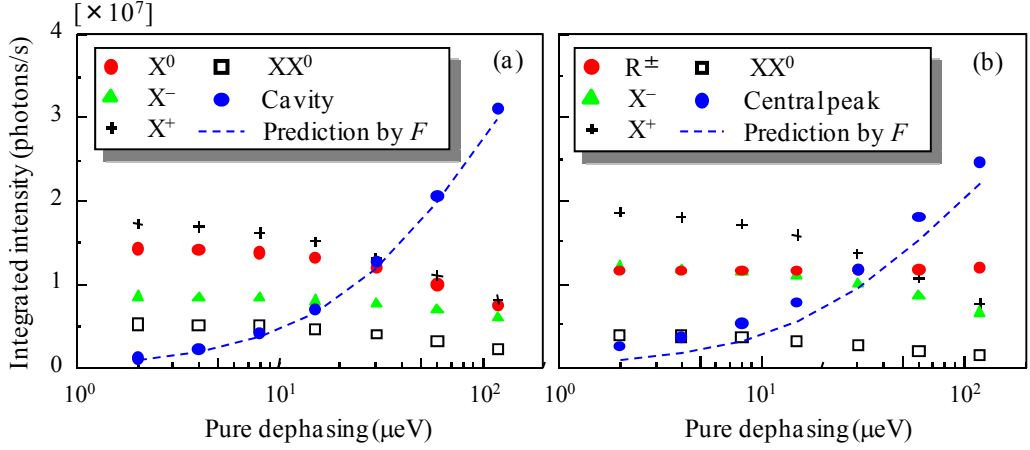

FIG. 3 (color). The dependence of integrated intensities on the pure-dephasing rate. The cavity is (a) +2.5 meV detuned from $BX^0$ and (b) tuned with $BX^0$, respectively. Each dashed line is calculated by $F$ with the integrated intensities of the optically active states ($BX^0$, $X^+$, $X^-$, and $XX^0$) in panel (a) and ($X^+$, $X^-$, and $XX^0$) in panel (b). Here, the AZE from $BX^0$ is not included in the prediction made using $F$ in panel (b), because $BX^0$ forms VRS states with the cavity and does not cause the AZE. $R^\pm$ represents the intensities of the upper (+) and lower (−) VRS states, and the central peak represents the additional bare cavity peak in VRS.

Unlike the TLS model, the present model allows various charging states. However, because the QD is allowed to take only one state at a time, each optically active state is expected to act as if a simple TLS is interacting with the cavity. Therefore, the intensity of the bare cavity line might be predicted to be a summation of the AZE from each detuned state. This idea is examined by comparing the integrated intensity of the bare cavity with the total AZE predicted by $F$ in Eq. (1) as a function of the $2\gamma_{phase}$. The results are shown in Fig. 3(a) and (b) for cases in which the cavity is detuned from all emission lines and tuned with $BX^0$, respectively. As shown in Fig. 3(a), the prediction made by $F$ is in good agreement with the direct calculation. This confirms that the off-resonant cavity light emission is attributable to the AZE from each detuned state. Moreover, the intensity of the bare cavity line in the triplet is also dominated by the AZE from the detuned optically active states of $X^+$, $X^-$, and $XX^0$, as shown in Fig. 3(b). This result directly shows that the central cavity peak in the triplet is provided from these detuned states with the help of the AZE. Therefore, the process forming the triplet can be explained as follows. When $BX^0$ is generated by the random carrier injection, the interaction with the cavity results in VRS. By contrast, when the other optically active states of $X^+$, $X^-$, and $XX^0$ are generated, the cavity photon can be created by the AZE, and at the same time the optically active states change into single hole, single electron, and $BX^0$ states, respectively. In this case, the cavity photon cannot form VRS states because they require the ground state of the QD together with a single photon in the cavity. As a result, the bare cavity lines appear in the spectra. Thus, VRS and the bare cavity lines are generated randomly, and the stochastically averaged spectrum results in the triplet.

Although the triplet is explained as described above, a small difference remains between the direct calculation and the prediction made using $F$, which suggests the existence of processes other than the AZE. This effect might arise from higher order processes, in which additionally injected carriers turn $BX^0$ into the other states while it still interacts with the cavity. This process could occur without the AZE (Appendix B). However, this effect is small under low-excitation conditions (for example, Fig. 2(a)), which is also evidenced by the small size of the deviation from the prediction made using $F$. Therefore, we conclude that the AZE, as discussed above, is the main process forming the triplet.

As the origin of the on-resonant triplet has been clarified, we now discuss the requirement to observe the triplet. From the mechanism described above, it is clear that that the additional cavity peak has to appear between the two VRS peaks, the magnitude of which is determined by the coupling constant of $g$. The coupling constant also determines the intensity of the central cavity peak in the triplet through the AZE. Therefore, the spectral triplet can be seen only when the coupling constant is sufficiently large. When this is not the case, the additional cavity peak becomes invisible behind the VRS peaks. This picture is approximate and incomplete, because the practical conditions needed to observe the triplet also depend on parameters such as the detuning of individual charging states, the pure-dephasing rate, and the optical damping rate of the cavity. However, the coupling constant is one of the most important conditions, which is consistent with the fact that the triplet is observed when the QD is deterministically embedded at the centre of the nanocavity [4,7,8], whereas the usual VRS [5,20,21] is observed when the position of the QD is not controlled.

### B. Cavity-mediated mixing

Thus, the on-resonant triplet, as well as the off-resonant cavity light emission, is explained well by the AZE from the detuned states. However, a difference remains between the calculation shown in Fig. 2(b) and the experimental findings [7]: that is, the cavity-mediated mixing of the $BX^0$ and $DX^0$ states. The spin-flip processes caused by the HF interactions are strong candidates for this phenomenon from an experimental viewpoint. Therefore, we now discuss the effect of HF interactions, which can be modeled by a nuclear magnetic field $\boldsymbol{B}_N = (B_{N,x}, B_{N,y}, B_{N,z})$ called the Overhauser field [22]. This field does not affect the heavy-hole pseudo-spins, so the interaction is described by $H_{\mathrm{hf}} = g_e \mu_B (B_{N,x} s_e^x + B_{N,y} s_e^y + B_{N,z} s_e^z)$ with $s_e^x \equiv (s_e^+ + s_e^-)/2$ and $s_e^y \equiv -\mathrm{i}(s_e^+ - s_e^-)/2$, where $g_e$ is the

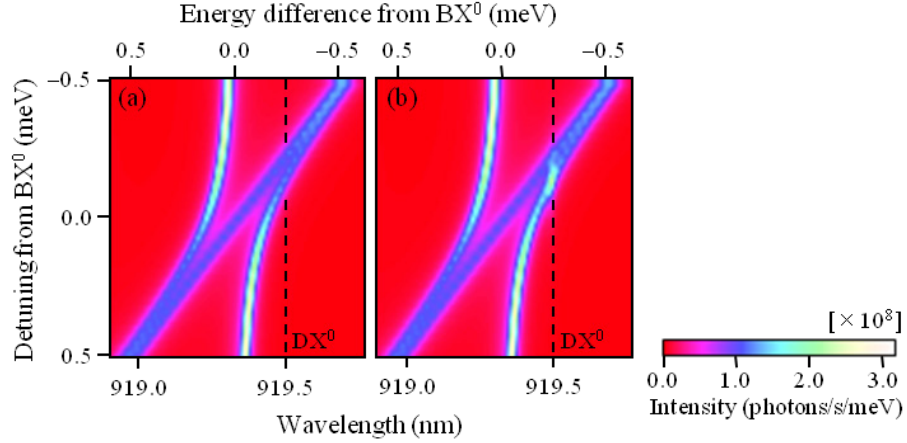

FIG. 4 (color). Calculated spectra around the BX$^0$ line (a) without and (b) with the Overhauser field with an equal magnitude of 20 mT in the $x$, $y$, and $z$ directions. The dashed line is the energy level of DX$^0$. The spectra for the other values of detuning are the same as in Fig. 2(b).

electron g-factor and $\mu_B$ is the Bohr magneton. The Overhauser field is temporally fluctuating in a precise sense, but the correlation time is relatively long (~1 ms) and so can be treated as a stochastically-dispersed quasi-static field [22]. In the case of InAs/GaAs QDs, the root–mean–square values for $B_{N,x}$, $B_{N,y}$, and $B_{N,z}$ are ~9.5–30 mT. Consequently, the spectra are calculated by adding $H_{hf}$ to $H_S(t)$ for a particular case of the Overhauser field with an equal magnitude of 20 mT in the $x$, $y$, and $z$ directions. As shown in Fig. 4, an additional line becomes activated when the Overhauser field is taken into account, corresponding to the cavity-mediated mixing of BX$^0$ and DX$^0$. Note that the calculated results are now in good agreement with the experiments [7]. The role of the Overhauser field is explained as follows. The Overhauser field in arbitrary directions breaks the symmetry of $H_S(t)$, and DX$^0$ is mixed with BX$^0$ [23,24]. Therefore, DX$^0$ becomes able to interact with photons. However, the Overhauser field is not sufficient to induce non-negligible light emission at DX$^0$ energy because of the considerable detuning between $DX^0$ and $BX^0$ (~0.25 meV) that makes the mixing negligible. For the same reason, $BX^0$ still interacts with photons strongly, and forms VRS states with the cavity when the detuning between the cavity and $BX^0$ becomes smaller. In this situation, if one of the VRS states is tuned to $DX^0$, the Overhauser field-induced mixing between the $DX^0$ and $BX^0$ components in the VRS states becomes strong because of their resonance. Therefore, the light emission is enhanced when $DX^0$ is tuned to the VRS state rather than the bare cavity. The hypothesis proposed in Ref. [7] is thus verified theoretically (for further discussions, see Appendix C).

## SUMMARY AND CONCLUSION

We have presented a comprehensive theory of couplings between a nanocavity and exciton complexes in a QD that allows various charge configurations. The numerically calculated results not only reproduce the unique features of the QD exciton complexes, but also successfully predict the on-resonant spectral triplet, which puzzled the community for several years. By comparing the intensity of the central peak in the triplet with the values predicted by the factor $F$, we have concluded that the on-resonant triplet is explained well by the AZE from the detuned states under a low-excitation regime.

Furthermore, the cavity-mediated mixing of bright and dark exciton states observed in recent experiments has been also studied based on our formulation, and it was demonstrated that this phenomenon can be reproduced by the effect of the Overhauser fields. This confirms that the cavity-mediated mixing is attributed to HF interactions theoretically.

The theory presented here can therefore fully and consistently explain the recent experimental results. Our formalism should prove to be useful for future studies, such as practical analyses of QIP using QD spins and cavity QED [1], design and optimization studies of the performance of single-photon sources [2], and numerical tests for new proposals of applications (for example, using the AZE) [10]. We believe that our findings will accelerate the understanding of the physical nature of solid-state cavity QED systems.

## ACKNOWLEDGMENT

This work was supported by Research Programs (Grant-in-Aid, Centre of Excellence, and Special Coordination Fund) for Scientific Research from the Ministry of Education, Culture, Sports, Science and Technology of Japan.

## APPENDIX A: CIRCULARLY-POLARIZED COMPONENTS

Here, we describe the two circularly-polarized components in $S_{\text{spon}}(\omega)$ for convenience. Because similar formalism to that used in the main text can easily be applied to these components, only the result is described, as follows:

$$S_{\text{spon}}(\omega) = S_{\text{L}}(\omega) + S_{\text{R}}(\omega), \tag{A1}$$

with

$$S_{\text{L}}(\omega) = \frac{2\Gamma_{\text{spon}}}{\pi} \text{Re}\left[\int_0^\infty d\tau \left\langle c_\uparrow^\dagger(t) d_\downarrow^\dagger(t) d_\downarrow(t+\tau) c_\uparrow(t+\tau) \right\rangle \exp(i\omega\tau - \gamma_{\text{reso}}\tau)\right], \tag{A2}$$

$$S_{\text{R}}(\omega) = \frac{2\Gamma_{\text{spon}}}{\pi} \text{Re}\left[\int_0^\infty d\tau \left\langle c_\downarrow^\dagger(t) d_\uparrow^\dagger(t) d_\uparrow(t+\tau) c_\downarrow(t+\tau) \right\rangle \exp(i\omega\tau - \gamma_{\text{reso}}\tau)\right], \tag{A3}$$

where $S_{\text{L}}(\omega)$ and $S_{\text{R}}(\omega)$ denote the left and right circularly-polarized components in $S_{\text{spon}}(\omega)$, respectively.

## APPENDIX B: HIGHER ORDER PROCESSES

In this section, we discuss the small difference between the prediction made using $F$ and the direct calculation shown in Fig. 3(b). In order to identify the physical reason for this occurrence, the injection rate-dependence of $S_{cav}(\omega) + S_y(\omega)$ around the $BX^0$ line is studied without pure-dephasing, as shown in Fig. 5 ($2\gamma_{reso}$ = 30 µeV). Here, $S_x(\omega)$ is not included in the spectra in order to eliminate the confusing $x$-polarized line. When the injection rate is relatively small (Fig. 5(a) and (b)), only the VRS can be seen. However, the additional peak appears again between the VRS when the injection rate becomes larger (Fig. 5(c)). In these cases, there is no AZE because the pure-dephasing is ignored; higher order processes should therefore become important, because there is a good chance that additional carriers can be injected into the $S$-shells while it still drives the cavity. The small difference shown in Fig. 3(b) might also arise from these higher order processes. However, as shown in Fig. 5(a) and (b), this effect is too small to form the triplet in a low-excitation regime. As a result, we conclude that the central peak in the triplet shown in Fig. 2(b) is dominated by the AZE.

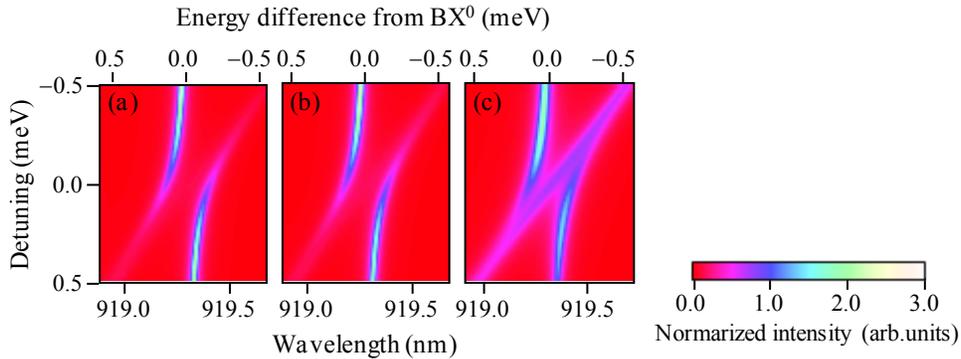

FIG. 5 (color). The injection rate-dependence of $S_{cav}(\omega) + S_y(\omega)$ around the $BX^0$ line without pure-dephasing. The injection rates are (a) $3.3\times10^1$ neV, (b) $3.3\times10^2$ neV, and (c) $3.3\times10^3$ neV, respectively. Each color scale is normalized in order to compare the spectral shapes.

## APPENDIX C: BEHAVIOR OF DARK EXCITONS

In the main text, we describe how symmetry breaking induced by the HF interactions makes the $DX^0$ states illuminant due to mixing with $BX^0$, which determines the behavior of the $DX^0$ states. However, it is also important to investigate whether other possibilities (such as strain-induced piezoelectric fields) rather than the HF interactions activate the $DX^0$ states. Therefore, here we discuss the optical selection rules for the $DX^0$ states from the viewpoint of the symmetry of the QD-confinement potential.

When the QD-confinement symmetry belongs to $D_{2d}$ point group symmetry (for example, a cylindrical QD), the exciton states consisting of an electron in the conduction band ($\Gamma_6$) and a heavy hole in the valence band ($\Gamma_6$) transform according to $\Gamma_6 \otimes \Gamma_6 = \Gamma_1 + \Gamma_2 + \Gamma_5$ in the double group notation [25]. Following a standard method to obtain selection rules in group theory, the $\Gamma_5$ is a doublet that is optically active for the light fields polarized in the $xy$ plane, which correspond to $BX^0$ states. By contrast, the $\Gamma_1$ and $\Gamma_2$ states are optically-inactive singlet states or $DX^0$ states. However, in a practical QD, the confinement potential belongs to $C_{2v}$ symmetry. When lowering the symmetry from $D_{2d}$ to $C_{2v}$, the $\Gamma_5$ state splits into two states, $\Gamma_5 \rightarrow \Gamma_2 + \Gamma_4$, which are optically active for the light fields polarized in the $x$ and $y$ direction, respectively. In a similar manner, $\Gamma_2 \rightarrow \Gamma_3$ and $\Gamma_1 \rightarrow \Gamma_1$ are obtained for $DX^0$ states. $\Gamma_3$ is still optically inactive, whereas $\Gamma_1$ becomes optically active for the light fields polarized in the $z$ direction.

This result suggests that one of the $DX^0$ states in the $C_{2v}$ confinement potential can interact with the $z$-polarized light fields. Therefore, it is possible for the $DX^0$ state to interact with the cavity field, which might lead to a similar consequence as shown in Fig. 3(b) when there are $z$-polarized components in the cavity at the QD position. In the case of a 2DPC nanocavity [11], however, there is no $z$ component at the centre of the 2DPC slab, because its structure has a mirror plane at the centre of the slab. Thus, at least in principle, the $DX^0$ states in the $C_{2v}$ confinement potential cannot interact with the 2DPC nanocavity without HF interactions when the QD is embedded at the centre of the slab.